\documentclass{osa-article}

\journal{osajournal}

\articletype{Research Article}

\begin{document}

\title{Single camera high repetition rate two-color formaldehyde planar laser-induced fluorescence thermometry with a wavelength-switching burst mode laser}

\author{Xunchen Liu,\authormark{1*} Yayao Wang,\authormark{1} Zhen Wang,\authormark{1} and Fei Qi,\authormark{1}}
\address{\authormark{1}School of Mechanical Engineering, Shanghai Jiao Tong University, 800 DongChuan Rd., Shanghai, China}
\email{\authormark{*}liuxunchen@sjtu.edu.cn} %% email address is required

\begin{abstract}
We consider a two-color formaldehyde PLIF thermometry scheme using a wavelength-switching injection seeding Nd:YAG laser at 355\,nm.
The 28183.5\,cm$^{-1}$ and 28184.5\,cm$^{-1}$ peaks of formaldehyde are used to measure low temperature combustion zone.
Using a burst mode amplifier and a high speed camera, high-repetition rate (20\,kHz) temperature field measurement is validated on a laminar coflow diffusion flame and demonstrated on a turbulent confined jet in hot crossflow flame. 
\end{abstract}

Flame temperature indicates thermodynamic and chemical reaction kinetics of the combustion species, determines thermal efficiency, and controls NOx emission of engines.
\textit{In-situ} measurement of the temperature field in combustion environment with high spatial and temporal resolution is one of the most difficult task of combustion diagnostic.
Commonly used non-contact laser-based diagnostic methods include planar Rayleigh-Brillouin scattering that imaging the gas density and spectroscopic methods such as line-of-sight absorption spectroscopy, planar laser induced fluorescence spectroscopy (PLIF), or pointwise coherent anti-stokes Raman spectroscopy that probing the Boltzmann distribution of gas phase molecules.

As a highly sensitive diagnostic method, two-line PLIF is particularly suitable to investigate the instantaneous temperature field of turbulent combustion in optical model combustors. %, because interference from window surface and soot particulate preclude the application of Rayleigh and Raman scattering.
Widely used OH-PLIF at 280-290 nm can investigate the high-temperature zone of flames after the flame front.
Using a dye laser with repetition rate at 10-20 Hz, the single pulse energy used for OH-thermometry is around 10-20 mJ\cite{dulin_assessment_2021,grib_two-dimensional_2021}.
High repetition-rate PLIF thermometry technique is highly desirable to study transient combustion instability, flame ignition, propagation, and blowout.
Recently, high-repetition rate PLIF measurement using burst mode laser as the pump source and injection-seeded optical parametric oscillator (OPO) has achieved 10 kHz repetition rate and single pulse energy around 5 mJ\cite{hsu_10_2021}.
However, due to low volume fraction of OH radical, the temperature of the low temperature region measured using two-line OH-PLIF thermometry would resulted in erroneous temperatuer\cite{chrystie_temperature_2013}. %At the mean time, two-line PLIF thermometry using such OPO output beam is still difficult to obtain due to the limited laser power.

In this letter, we propose a new two-line PLIF thermometry scheme that directly uses the third harmonic of Nd:YAG lasers to probe formaldehyde intermediate from hydrocarbon fuel low temperature pre-heat zone of combustion.
Formaldehyde is widely present during the ignition and cool flame stage of combustion and utilized to mark the unsteady combustion that drives instability\cite{fugger_structure_2019} and low-temperature zone of lift-off non-premixed flames\cite{macfarlane_stabilisation_2017}.
We demonstrate the two-line formaldehyde PLIF thermometry at 20 kHz, measuring two dimensional low-temperature region of a coflow diffusion flame and a turbulent jet in hot crossflow flame are measured using a single CMOS camera and a wavelength-switching burst mode laser.

Formaldehyde is a prototype asymmetric top polyatomic molecule, with its spectroscopy thoroughly studied by both experimental and theoretical techniques.
Among several formaldehyde-PLIF excitation schemes using the $\tilde{A}^1A_2-\tilde{X}^1A_1$ system in the UV region that include out-of-plane bending vibration mode (v4) and the carbonyl stretching mode (v2), weak absorption at 355\,nm covers the $b$-type transitions of the $2^0_04^1_0$ band of formaldehyde\cite{parkin_band_1965}. 
The 355 nm excitation directly using third harmonic of a Nd:YAG laser has the main advantage of, especially at high-repetition rate, large excitation energy in field applications. 
It also avoids major interference species and various photo-dissociative channels of formaldehyde that leads to radicals\cite{moore_formaldehyde_1983}.
However, very little high resolution work has been done in this region.
Brackmann et al.\cite{brackmann_strategies_2005} and Gabet et al.\cite{gabet_narrowband_2014} showed that narrow band excitation at the formaldehyde absorption peaks around 28182.5\,cm$^{-1}$ and 28183.5\,cm$^{-1}$ in DME flames using the third-harmonic of injection-seeded Nd:YAG lasers avoids major interference from additional species such as PAH and soot, comparing to broadband excitation.
Rotationally resolved UV absorption cross sections from 351-356\,nm have been measured using FTIR at room temperature and low pressure\cite{co_rotationally_2005}.

\begin{figure}[h]
{
\includegraphics[width=0.95\textwidth]{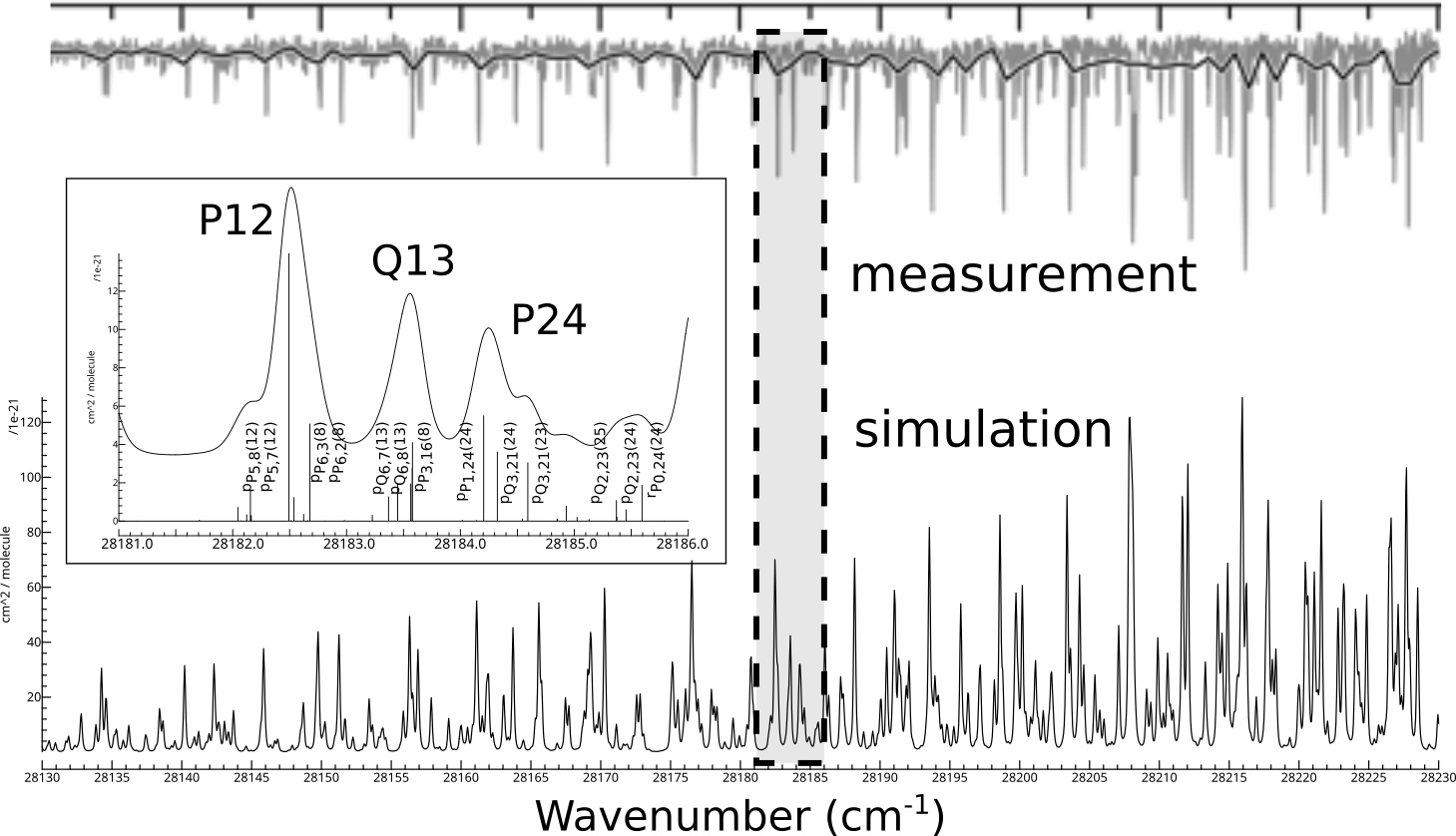}
}
\caption
{Assignment and simulation of the experimental absorption cross sections\cite{co_rotationally_2005} of formaldehyde at room temperature.}
\label{fig-simulation-spec}   
\end{figure}

Figure\,\ref{fig-simulation-spec} shows simulation of the UV absorption spectra of formaldehyde in the 28180-28190\,cm$^{-1}$ region using ground state rotational constants taken from microwave-optical transitions\cite{winnewisser_determination_1979} and excited state constants from Ramsay and Till\cite{clouthier_spectroscopy_1983}.
The spin-statistical weight for rotational levels with even/odd $K_a$ due to equivalent hydrogen nuclei is 1:3. 
Overall, the simulation can match the FTIR measurement satisfactorily, with small discrepancy of high $J$ transition intensities due to perturbation from other vibrational states.
The insert of Fig.\,\ref{fig-simulation-spec} shows the region accessible by Nd:YAG lasers, including the 28182.5\,cm$^{-1}$ peak marked as the P12, mainly consisting of the $^{\Delta K}\Delta J_{K_aK_c}(J)=^pQ_{5,7/8}(12)$ transitions.
Two peaks at 28183.5\,cm$^{-1}$ and 28184.2\,cm$^{-1}$, corresponding to $^pQ_{6,7/8}(13)$ and $^pP/Q_{1/3,24/21}(24)$ transitions have large lower state energy difference.
Marked as Q13 and P24, they provide good line-pair candidate for two-line formaldehyde PLIF thermometry in the following experimental study. 

\begin{figure}[h]
{
\includegraphics[width=0.95\textwidth]{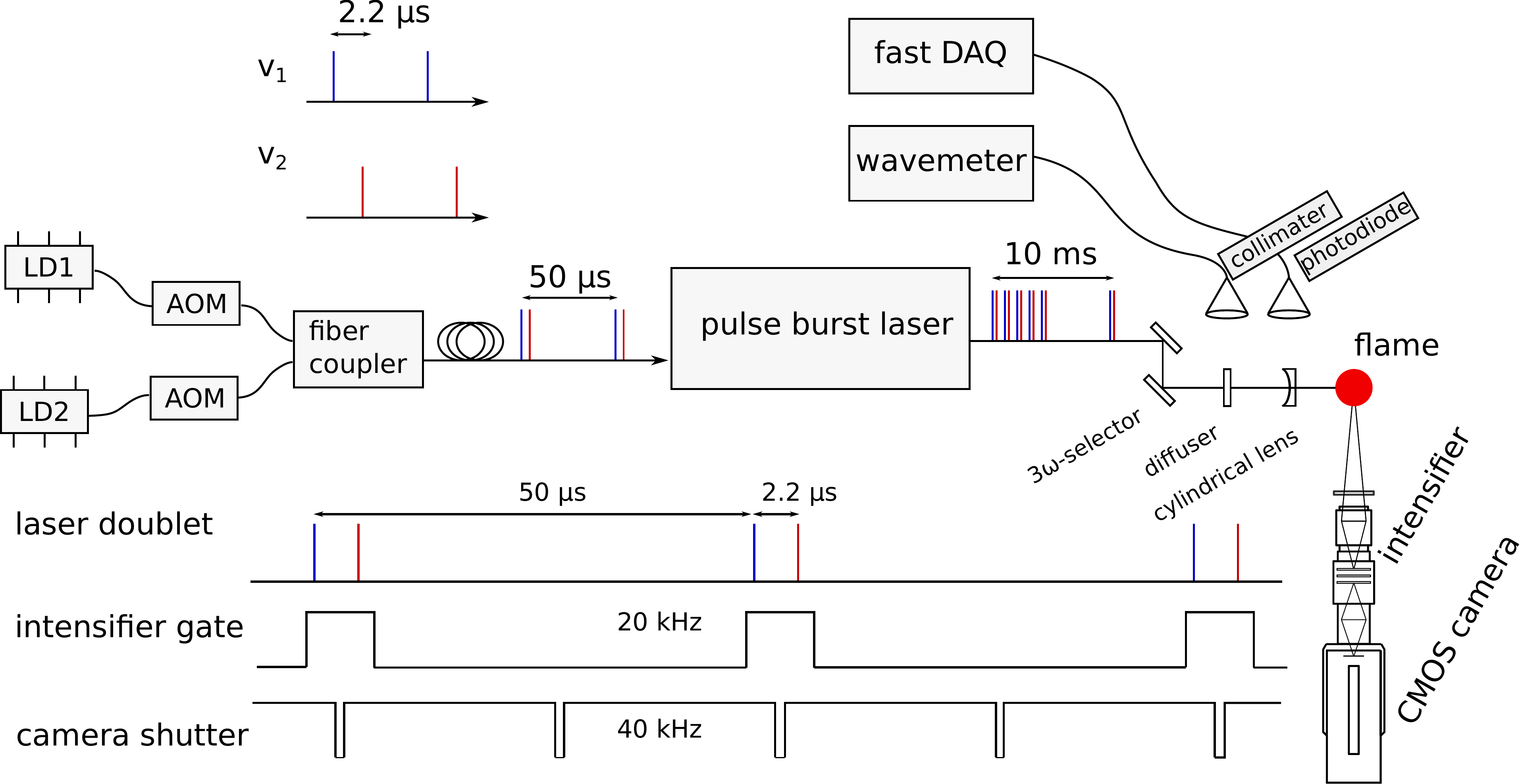}
}
\caption
{Experimental setup of single camera wavelength-switching two-line PLIF using a burst mode laser}
\label{fig-setup}   
\end{figure}

The schematic experimental setup of the two-line PLIF measurement using a wavelength-switching burst mode laser is shown in Figure\,\ref{fig-setup}.
The seed of the MOPA laser is 20\,kHz rate 10\,ns width doublet pulses from two fiber coupled and AOM chopped laser diodes with 2.2\,$\mu$s separation, each tuned to different absorption wavelength of formaldehyde.
The doublet train is injected to a flashlamp-pumped Nd:YAG burst mode amplifier (QuasiModo 1000, Spectral Energies) and frequency tripled to produce narrowband (0.01\,cm$^{-1}$) third-harmonic with single pulse energy around 50 mJ.
The two laser wavelength of the 3$\omega$ doublets are tuned to 28183.5\,cm$^{-1}$ (Q13) and 28184.2\,cm$^{-1}$ (P24) by adjusting the temperature and current of two laser diodes, monitored by a calibrated wavemeter (HighFiness, WS6-200-UV).
The 355\,nm output is then formed to a light sheet using a diffuser and two cylindrical lenses. 
The CH$_2$O fluorescence band is captured with a 376-478\,nm hendeca-band bandpass filter (FF01-CH$_2$O, Semrock) and imaged using an image intensifier (HiCATT 25 1:1, Lambert) and a high speed CMOS camera (Photron, SA-Z).
The intensifier operates at 20\,kHz synchronized to the laser doublet while the CMOS camera operates at 40\,kHz in frame-straddling mode. % to capture the intensifier amplified images. 
In this way, two-line PLIF image can be captured using a single CMOS camera. The  2.2\,$\mu$s separation of the doublet is required due to shutter slope. In this experiment, with pixel size $\sim$2$\mu$m and flow speed below 100\,m/s, the separation does not introduce measurement error. Reducing this separation at the cost of SNR for higher flow speed is also possible.

\begin{figure}[h]
{
\includegraphics[width=0.95\textwidth]{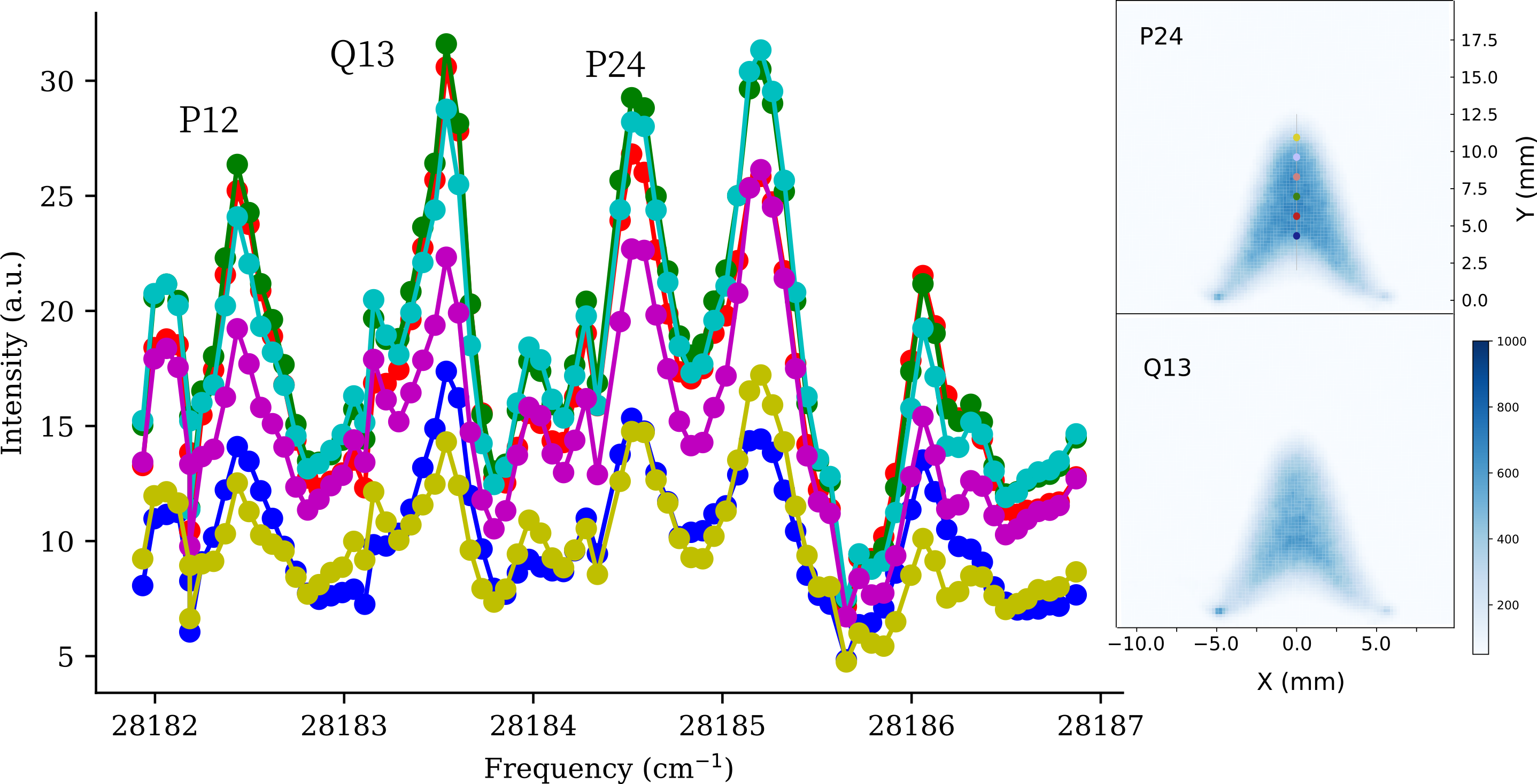}
}
\caption
{Scanned total LIF spectra of formaldehyde in laminar coflow diffusion flame.}
\label{fig-coflow-spec}   
\end{figure}

To validate the two-line formaldehyde-PLIF thermometry technique we proposed, 
we first measured the formaldehyde fluorescence spectra and temperature field of a coflow DME diffusion flame.
The canonical laminar diffusion flame is generated by a Santoro type coannular burner with 10\,mm inner diameter exit for fuel ($X_{DME}=30\%$, $X_{N2}=70\%$, 0.55\,m/s) and 86\,mm diameter of coflow air\cite{liu_two-dimensional_2018}.
The coflow diffusion flame has high formaldehyde concentration due to thermal decomposition of DME in the cool flame region, similar to DME Bunsen flame measurement\cite{brackmann_laser-induced_2003,brackmann_strategies_2005}.
Formaldehyde PLIF images measured at the P24 and Q13 peak of the coflow flame are shown in Figure\,\ref{fig-coflow-spec}.
By scanning the third harmonic wavelength and recording the scatterd laser intensity using a photodiode for correction, we can obtain the laser power corrected total fluorescence spectra of each point of the laminar coflow flame from the PLIF images.
Figure\,\ref{fig-coflow-spec} are the LIF spectra of formaldehyde at the different height of the coflow flame, which can confirm the previous findings that tuning within one wavenumber of the 355\,nm laser wavelength can generate five to six fold of PLIF signal intensity difference due to the absorption structure\cite{bai_absorption_2004}.
The LIF spectra can be easily tracked to the simulated absorption spectrum with the P12, Q13, and P24 peaks clearly assigned, indicating that the rotational state dependent effects on upper states dynamics, quenching and broadening of the LIF signal are similar for these peaks due to the small value of the rotational constants.
Therefore, the simulated absorption ratio can be used to calculate the flame temperature from ratio of fluorescence intensities. 

\begin{figure}[h]
{
\includegraphics[width=0.95\textwidth]{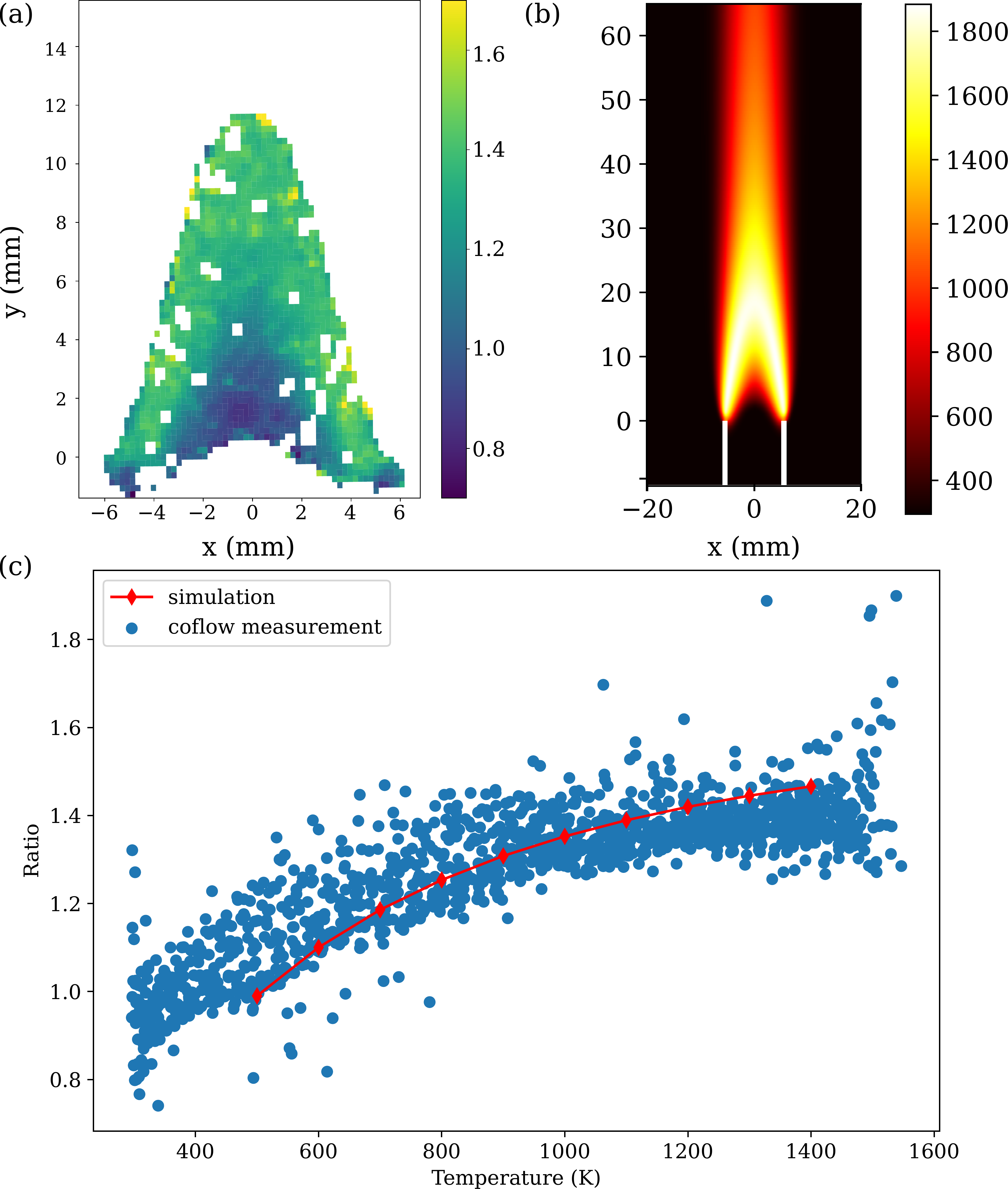}
}
\caption
{Validation of the two-line PLIF ratio and temperature relationship from spectral simulation in the laminar coflow diffusion flame. (a) experimentally measured PLIF signal ratio; (b) CFD simulation of coflow flame temperature field; (c) correlation between the PLIF ratio from (a) and CFD simulated temperature, compared with absorption spectra simulation. }
\label{fig-coflow-scatter}   
\end{figure}

By taking the intensity ratio of P24 over Q13 PLIF measurement shown in Figure\,\ref{fig-coflow-spec} and mapping the ratio distribution shown in Figure\,\ref{fig-coflow-scatter}(a) to the CFD calculated temperature field shown in Figure\,\ref{fig-coflow-scatter}(b), we can obtain the scatter plot of the fluorescence signal ratio to the flame temperature shown in Figure\,\ref{fig-coflow-scatter}(c). 
The experimental measured ratio-to-temperature function shows excellent agreement with the simulated absorption peak area ratio between Q13 and P24.
Thus, the canonical laminar flame temperature measurement validates the 355\,nm formaldehyde PLIF thermometry method proposed in this paper and gives an estimation of the accuracy within 100 K, without consideration of laser line width and imaging collection factors\cite{kostka_comparison_2009}.
We should also mention that the strong LIF peak at 28185.2\,cm$^{-1}$ shown in Figure\,\ref{fig-coflow-spec} cannot be confidently assigned to formaldehyde, although it appears in accordance with other formaldehyde peaks in the elevated temperature region and serve as a nice choice of line pair for thermometry measurement.
Further dispersive LIF spectrum measurement is need to eliminate potential PAH and soot interference.

\begin{figure}[h]
{
\includegraphics[width=0.95\textwidth]{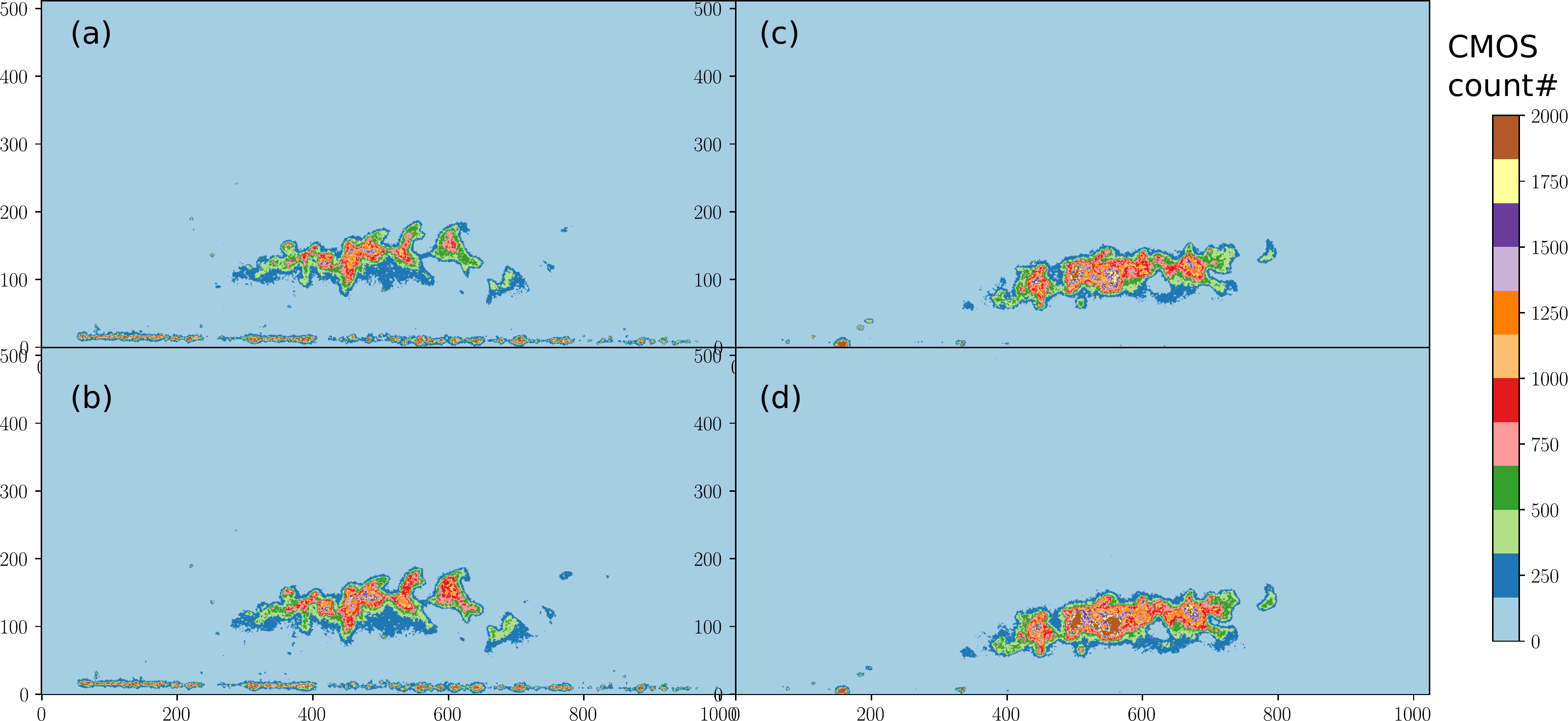}
}
\caption
{Instantaneous PLIF doublet image of the jet in hot crossflow flame with (a)/(b) round flat nozzle and (c)/(d) slotted nozzle with inclining angle}  
\label{fig-crossflow-plif}   
\end{figure}

\begin{figure}[h]
{
\includegraphics[width=0.95\textwidth]{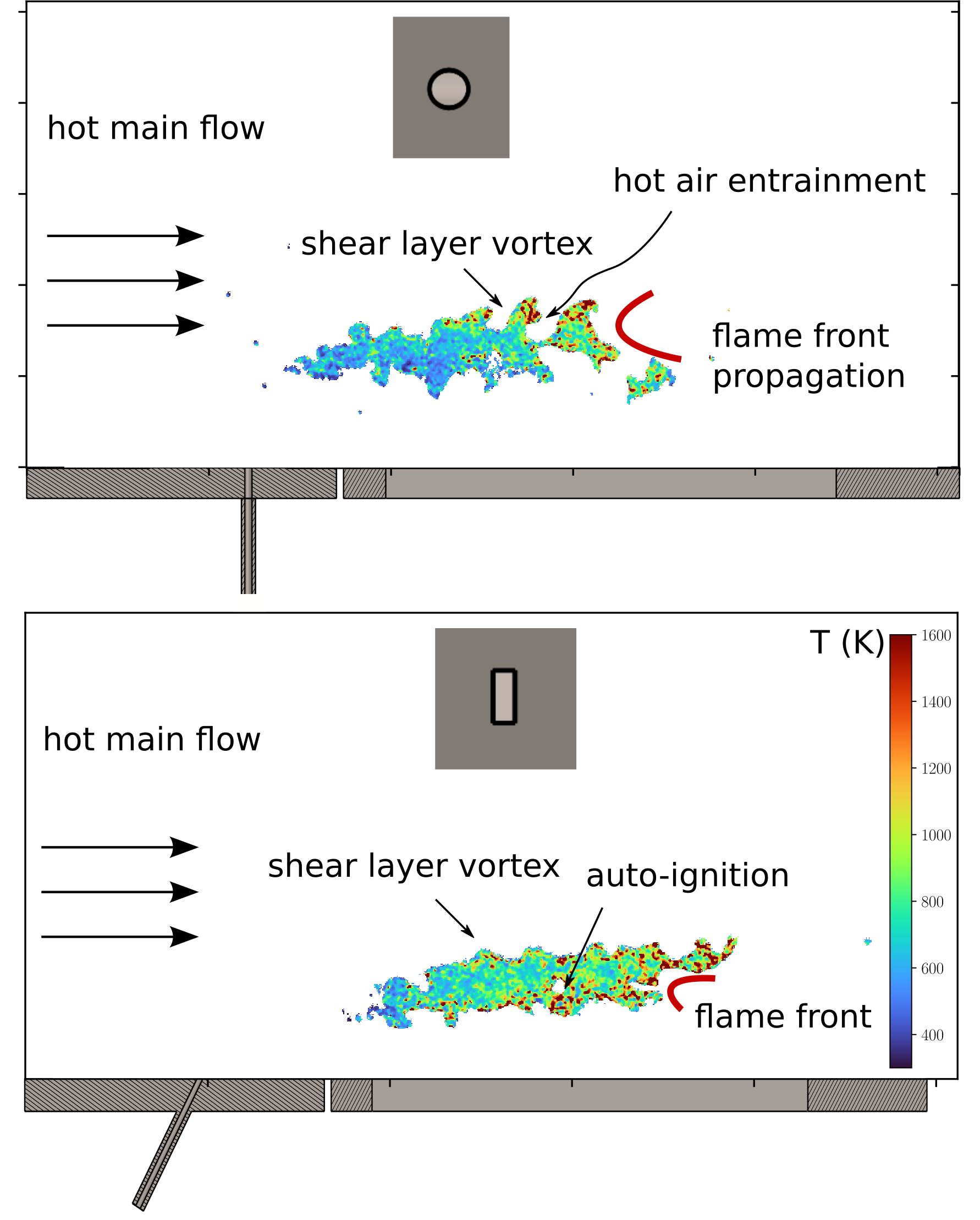}
}
\caption
{ Instantaneous temperature field showing the low-temperature pre-ignition region of the jet in hot crossflow flame before the flame front. (a) round flat nozzle; (b) slotted nozzle with inclining angle.}
\label{fig-crossflow-T}   
\end{figure}

The coflow flame measurement is similar to other line-scanning PLIF thermometry to probe steady flames\cite{zelenak_demonstration_2019,wang_quantitative_2019}.
Here, we demonstrate the advantage of high-repetition rate wavelength-switching PLIF thermometry by interrogation of a turbulent jet in hot crossflow flame. 
The burner consists of a first stage perforated plate flat flame with lean burn methane/air mixer, a contraction section and a second stage jet in hot crossflow flame with DME fuel.
The main flow after the contraction section is $\sim$1600\,K 10\,m/s and the jet is 300\,K 50\,m/s. 
We examine the effects of nozzle geometry by using a 2\,mm diameter round flat nozzle with normal injection angle and a 3$\times$1\,mm slotted flat nozzle with $30^{\circ}$ inclining angle.
Typical instantaneous PLIF measurement and temperature field from the PLIF intensity ratio of P24/Q13 doublet are shown in Figure\,\ref{fig-crossflow-plif} and Figure\,\ref{fig-crossflow-T}.

In Figure\,\ref{fig-crossflow-T}(a), with a round flat nozzle, the formaldehyde PLIF thermometry measurement clearly reveals the temperature increase from around $\sim$500\,K to $\sim$1600\,K along the jet trajectory, agree with laminar flame calculation\cite{gabet_comparison_2013}.
The temperature raise of the jet is due to mixing with the surrounding hot air, mainly through shear layer vortices and hot air entrainment from the main flow, which also brings the equivalence ratio of the fuel-air mixture close to stoichiometric, leading to ignition of the mixture at the flame front through flame propagation.
The experimental measurement shows that the high temperature region is mainly at the windward-side of the jet.
The temperature field measurement of the slotted flat nozzle with $30^{\circ}$ inclining angle is shown in Figure\,\ref{fig-crossflow-T}(b).
With smaller Kelvin–Helmholtz vortices, the temperature increase along the jet trajectory is quicker due to larger perimeter compared to the round nozzle, leading to closer and more frequent auto-ignition of the fuel-air mixture when the temperature reaches around 1600\,K. %, in contrary to the partially premixed flame propagation case in the round nozzle case. 
We can also observe that the high temperature region appears at both windward and lee-side of the jet, probably due to the recirculation zone.
The enhanced mixing effects of the slotted nozzle, revealed by the temperature field measurement, contribute to enhancement of flame stability\cite{pinchak_effects_2018}.

\begin{figure}[h]
{
\includegraphics[width=0.95\textwidth]{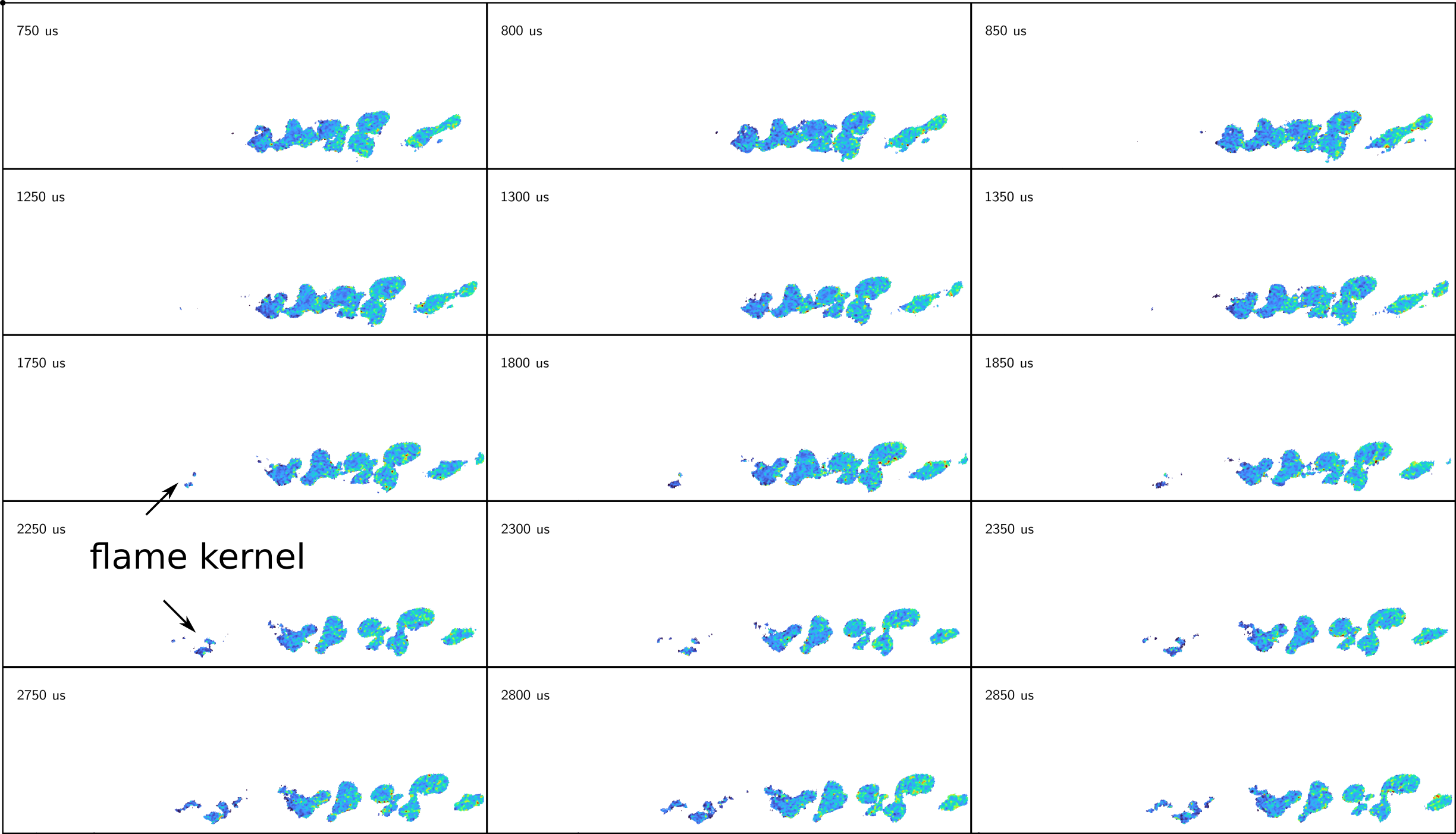}
}
\caption
{Temporal dynamics of the jet in hot crossflow temperature field with intermittent lift-off height. } 
\label{fig-crossflow-time}   
\end{figure}

In this letter, high-repetition PLIF thermometry measurement at 20\,kHz is enabled by using a burst mode amplifier.
Here, we examine the effects of crossflow temperature on the auto-ignition dynamics. % of the jet in hot crossflow flame.
The main flow temperature is reduced from $\sim$1600\,K to $\sim$1400\,K by varying the main flow equivalence ratio.
Flame base intermittency is observed since auto-ignition delay increases with temperature drop of the main flow and fuel air mixture.
Figure\,\ref{fig-crossflow-time} shows the temporal sequence of the formaldehyde PLIF temperature field with 50\,$\mu$s interval.
The temperature field dynamics of DME pyrolysis and flame kernel initiation/growth can be clearly observed.
The temperature increase along the trajectory is lower compared to main flow at $\sim$1600\,K , leading to a lower ignition temperature at the equivalence ratio controlled flame font.
The local temperature increase on a rate that is relatively unchanged during the auto-ignition even on Lagrangian viewpoint.
The effects of surrounding temperature on flame base stability has been observed on a jet in hot coflow flame with DME fuel\cite{macfarlane_stabilisation_2017}.
The current study provides crucial information of the flame base temperature and shed new insight of the auto-ignition mechanism.

In summary, we demonstrate high-repetition rate (20\,kHz) two-line formaldehyde PLIF thermometry of the low-temperature combustion region using a novel wavelength-switching scheme, a burst mode amplifier, and a single high-speed camera.  
High-repetition rate 2D temperature field measurements are obtained on both canonical laminar coflow diffusion flame and a jet in hot crossflow flame, in which the effects of nozzle geometry and crossflow temperature on the auto-ignition dynamics are revealed.
This work expands the capability of the widely used formaldehyde PLIF strategy to quantitative 2D temperature field measurement.  
The wavelength-switching method is easy to deploy and less complexity/expensive.
The advantage of using a single camera and intensifier avoids intensifier white field calibration and the error introduced by image overlapping.
The advantage of using a wavelength-switching doublet from single laser avoids laser intensity fluctuation and laser beam profile drift. 
These advantages make the current method very suitable to obtain high quality temperature field results for high-repetition combustion diagnostic.
Though demonstrated at 355\,nm, the wavelength-switching method can be easily tuned to other laser wavelength to performed CARS/PLIF-thermometry or LII/PLIF measurement, by coupling with optical parametric oscillators or dye lasers.
Although used with a burst mode amplifier, the wavelength-switching scheme can be used in any MOPA lasers for two-line based laser diagnostic.

\textbf{Funding.} National Natural Science Foundation of China (NSFC) (51606123, 91541201). 

\textbf{Acknowledgment.} The authors thank Dr. Hanfeng Jin and Dr. Junjun Guo for private communication of coflow flame CFD calculation.

\bibliography{xliu}

\end{document}